\documentstyle[12pt,epsf]{article}

\newcommand{\sect}[1]{\setcounter{equation}{0}\section{#1}}
\newcommand{\rf}[1]{(\ref{#1})}

\begin{document}

\title{Primordial Black Holes:
      \\ Tunnelling vs.\ No Boundary Proposal\thanks{Contribution to
        the Proceedings of COSMION 96 (Moscow, May 1996)}}
\author{{\sc Raphael Bousso}\thanks{\it R.Bousso@damtp.cam.ac.uk} \ 
  and {\sc Stephen W. Hawking}\thanks{\it S.W.Hawking@damtp.cam.ac.uk}
      \\[1 ex] {\it Department of Applied Mathematics and}
      \\ {\it Theoretical Physics}
      \\ {\it University of Cambridge}
      \\ {\it Silver Street, Cambridge CB3 9EW}
       }
\date{DAMTP/R-96/34}

\maketitle

\begin{abstract}

  In the inflationary era, black holes came into existence together
  with the universe through the quantum process of pair creation. We
  calculate the pair creation rate from the no boundary proposal for
  the wave function of the universe. Our results are physically
  sensible and fit in with other descriptions of pair creation. The
  tunnelling proposal, on the other hand, predicts a catastrophic
  instability of de~Sitter space to the nucleation of large black
  holes, and cannot be maintained.

\end{abstract}

\pagebreak

\sect{Introduction}

\subsection{Primordial Black Holes}

We now have good observational evidence for black holes from stellar
masses up to super-massive holes of $ 10^8 $ to $ 10^{10} $ solar
masses and maybe even more. However, one can also speculate on the
possible existence of black holes of much lower mass. These are the
holes for which quantum effects can be important. Such holes could not
form from the collapse of normal baryonic matter because degeneracy
pressure will support white dwarfs or neutron stars below the
Chandrasekhar limiting mass. One can express this limiting mass as $
m_{\rm Planck} (m_{\rm Planck}/m_{\rm baryon})^2 $. Its value is about
a solar mass, which might seem a coincidence, but there are good
anthropic principle reasons why stars should be just on the verge of
gravitational collapse.

This limiting mass applies only to the formation of black holes
through the gravitational collapse of fermions. In the case of bosons
the limiting mass is given by $ m_{\rm Planck} (m_{\rm Planck}/m_{\rm
  boson}) $. To form a black hole by the gravitational collapse of
bosons, they need to have a non-zero mass and either be stable or
have a fairly long life. About the only candidate is the axion, which
might have a mass of about $ 10^{-5} {\rm \/ eV} $. In this case the
limiting mass would be about the mass of the Earth, which is still
quite high, and too large for quantum effects to be observable. To get
black holes that are significantly smaller, one could not rely on
gravitational collapse, but would have to shoot matter together with
high energies.  John Wheeler once calculated that if one made a
hydrogen bomb with all the deuterium from the oceans, the centre would
implode so violently that a little black hole would be formed. Perhaps
fortunately, this experiment is unlikely to be performed. Thus the
only place where tiny black holes might be formed is the early
universe.

Previous discussions of black holes formed in the early universe have
concentrated on black holes formed by matter coming together during
the radiation era or first order phase transitions. Recent work on the
critical behaviour of gravitational collapse has shown it is possible
to form black holes in these situations. However, it is difficult
because one has to arrange for matter to be fired together at high
speed and accurately focused into a small region. Yet if too much
matter is fired together it forms a closed universe on its own, with
no connection with our universe. Such a separate universe would not be
a black hole in our universe.

Black holes formed by collapse, or by hurling matter together, are not
really primordial, in the sense that they do not form until a definite
time after the beginning of the universe. On the other hand, the black
holes we are going to consider form by the quantum process of pair
creation and are truly primordial, in that they can be considered to
have existed since the beginning of the universe.

\subsection{Inflation}

It is generally assumed that the universe began with a period of
exponential expansion called inflation. This era is characterised by
the presence of an effective cosmological constant $\Lambda_{\rm eff}$
due to the vacuum energy of a scalar field $\phi$. In chaotic
inflation~\cite{Lin83,Haw84} the effective cosmological constant
typically starts out large and then decreases slowly until inflation
ends when $\Lambda_{\rm eff} \approx 0$. Correspondingly, these models
predict cosmic density perturbations which are proportional to the
logarithm of the scale. On scales up to the current Hubble radius
$H_{\rm now}^{-1}$, this agrees well with observations of near scale
invariance. However, on much larger length scales of order $H_{\rm
  now}^{-1} \exp(10^5)$, perturbations are predicted to be on the
order of one. Of course, this means that the perturbational treatment
breaks down; but it indicates that black holes may be created.

Linde~\cite{Lin86a,Lin86b} noted that in the early stages of
inflation, when the strong density perturbations originate, the
quantum fluctuations of the inflaton field are much larger than its
classical decrease per Hubble time. He concluded that therefore there
would always be regions of the inflationary universe where the field
would grow, and so inflation would never end globally (``eternal
inflation''). However, this approach only allows for fluctuations of
the field. One should also consider fluctuations which change the
topology of space-time.  This topology change corresponds to the
formation of a pair of black holes. The pair creation rate can be
calculated using instanton methods, which are well suited to this
non-perturbative problem.

\subsection{Pair Creation}

Quantum pair creation is only possible on a background that provides a
force which pulls the pair apart.  In the case of a virtual
electron-positron pair, for example, the particles can only become
real if they appear in an external electric field. Otherwise they
would just fall back together and annihilate.  The same holds for
black holes; examples in the literature include their pair creation on
a cosmic string~\cite{HawRos95a}, where they are pulled apart by the
string tension; or the pair creation of magnetically charged black
holes on the background of Melvin's universe~\cite{Gib86}, where they
are separated by a magnetic field. In our case, the black holes will
be accelerated apart by the inflationary expansion of the universe.
While preventing classical gravitational collapse, this expansion
provides a suitable background for the quantum pair creation of black
holes.

After the end of inflation, during the radiation and matter dominated
eras, the effective cosmological constant was nearly zero. Thus the
only time when black hole pair creation was possible in our universe
was during the inflationary era, when $\Lambda_{\rm eff}$ was large.
Moreover, these black holes are unique since they can be so small that
quantum effects on their evolution are important. Indeed, their
evolution turns out to be quite interesting and
non-trivial~\cite{BouHaw96}. Here we will only describe the creation
of black holes, summarising a more rigourous treatment~\cite{BouHaw95}.
We focus on the consequences for the choice of the prescription for
the wave function of the universe.

In the standard semi-classical treatment of pair creation, one finds
two instantons: one for the background, and one for the objects to be
created on the background. From the instanton actions $I_{\rm bg}$
and $I_{\rm obj}$ one calculates the pair creation rate $\Gamma$:
\begin{equation}
\Gamma =
\exp \left[ - \left( I_{\rm obj} - I_{\rm bg} \right) \right],
\label{eq-pcr-usual}
\end{equation}
where we neglect a prefactor. This prescription has been very
successfully used by a number of authors
recently~\cite{DowGau94a,ManRos95,HawRos95b,CalCha96}
for the pair creation of black holes on various backgrounds. It is
motivated not only by analogies in quantum mechanics and quantum field
theory~\cite{Col77,CalCol77}, but also by considerations of black hole
entropy~\cite{GarGid94,DowGau94b,HawHor95}.

In this paper, however, we will obtain the pair creation rate through
a somewhat more fundamental procedure. Since we have a cosmological
background, we can apply the tools of quantum cosmology, and use the
wave function of the universe to describe black hole pair creation.
Two different prescriptions have been put forward for the calculation
of this wave function: Vilenkin's tunnelling proposal~\cite{Vil86}, and
the Hartle-Hawking no boundary proposal~\cite{HarHaw83} (reviewed in
Sec.~\ref{sec-nbp}). We will describe the creation of an inflationary
universe by a de~Sitter type gravitational instanton, which has the
topology of a four-sphere, $S^4$. In this picture, the universe
starts out with the spatial size of one Hubble volume. After one
Hubble time, its spatial volume will have increased by a factor of
$e^3 \approx 20$. However, by the de~Sitter no hair theorem, we can
regard each of these $20$ Hubble volumes as having been nucleated
independently through gravitational instantons. With this
interpretation, we are allowing for black hole pair creation, since
some of the new Hubble volumes might have been created through a
different type of instanton that has the topology $S^2 \times S^2$ and
thus represents a pair of black holes in de~Sitter
space~\cite{GinPer83}. Using the no boundary proposal, we assign
probability measures to both instanton types. We then estimate the
fraction of inflationary Hubble volumes containing a pair of black
holes by the fraction $\Gamma$ of the two probability measures. This
is equivalent to saying that $\Gamma$ is the pair creation rate of
black holes on a de~Sitter background.

In Sec.~\ref{sec-fixed-cc} we describe the relevant instantons and
calculate the pair creation rate.  The result is compared with that
obtained from the tunnelling proposal in Sec.~\ref{sec-discussion},
where we demonstrate that the usual description of pair creation,
Eq.~\rf{eq-pcr-usual}, arises naturally from the no boundary proposal.
We shall use units in which $m_{\rm P} = \hbar = c = k = 1 $.

\sect{The Wave Function of the Universe} \label{sec-nbp}

The prescription for the wave function of the universe has long been
one of the central, and arguably one of the most disputed issues in
quantum cosmology. The two competing proposals differ in their choice
of boundary conditions for the wave function.

\subsection{No Boundary Proposal} \label{ssec-nbp}

According to the no boundary proposal, the quantum state of the
universe is defined by path integrals over Euclidean metrics $ g_{\mu
  \nu} $ on compact manifolds $ M $. From this it follows that the
probability of finding a three-metric $ h_{ij} $ on a spacelike
surface $ \Sigma $ is given by a path integral over all $ g_{\mu \nu}
$ on $ M $ that agree with $ h_{ij} $ on $ \Sigma $. If the spacetime
is simply connected (which we shall assume), the surface $ \Sigma $
will divide $ M $ into two parts, $ M_+ $ and $ M_- $. One can then
factorise the probability of finding $ h_{ij} $ into a product of two
wave functions, $ \Psi_+ $ and $ \Psi_- $.  $ \Psi_+ $ ($\Psi_-$) is
given by a path integral over all metrics $ g_{\mu \nu} $ on the
half-manifold $ M_+ $ ($M_-$) which agree with $ h_{ij} $ on the
boundary $ \Sigma $. In most situations $ \Psi_+ $ equals $ \Psi_- $.
We shall therefore drop the suffixes and refer to $ \Psi $ as the wave
function of the universe. Under inclusion of matter fields, one
arrives at the following prescription:
\begin{equation}
\Psi[h_{ij}, \Phi_{\Sigma}] = 
\int \! D(g_{\mu\nu}, \Phi) \,
 \exp \left[ -I(g_{\mu\nu}, \Phi) \right],
 \label{eq-nbp}
\end{equation}
where $(h_{ij}, \Phi_{\Sigma})$ are the 3-metric and matter fields on
a spacelike boundary $\Sigma$ and the path integral is taken over all
compact Euclidean four geometries $g_{\mu\nu}$ that have $\Sigma$ as
their only boundary and matter field configurations $\Phi$ that are
regular on them; $I(g_{\mu\nu}, \Phi)$ is their action.  The
gravitational part of the action is given by
\begin{equation}
I_E = -\frac{1}{16\pi} \int_{M_+} \!\!\! d^4\!x\, g^{1/2}(R-2\Lambda) 
      -\frac{1}{8\pi}  \int_{\Sigma} \! d^3\!x\, h^{1/2} K,
\label{eq-action}
\end{equation}
where $R$ is the Ricci-scalar, $\Lambda$ is the cosmological constant,
and $K$ is the trace of $K_{ij}$, the second fundamental form of the
boundary $\Sigma$ in the metric $g$.

We shall calculate the wave function semi-classically, using a
saddle-point approximation to the path integral; and from the wave
function we shall calculate the pair creation rate. The method can be
outlined as follows. One is interested in two types of inflationary
universes: one with a pair of black holes, and one without. They are
characterised by spacelike sections of different topology. For each of
these two universes, one has to find a classical Euclidean solution to
the Einstein equations (an instanton), which can be analytically
continued to match a boundary $ \Sigma $ of the appropriate topology.
One then calculate the Euclidean actions $I$ of the two types of
saddle-point solutions.  Semiclassically, it follows from
Eq.~\rf{eq-nbp} that the wave function is given by
\begin{equation}
\Psi = \exp \left( -I \right),
\label{eq-nbp-sca}
\end{equation}
neglecting a prefactor.  One can thus assign a probability measure to
each type of universe:
\begin{equation}
P = \left| \Psi \right|^2 = \exp \left( -2I^{\rm Re} \right),
\label{eq-prob-measure}
\end{equation}
where the superscript `Re' denotes the real part.  As explained in the
introduction, the ratio of the two probability measures gives the rate
of black hole pair creation on an inflationary background, $\Gamma$.

The probability measure $P$ for the nucleation of a space-time should
be proportional to the number of possible quantum states it contains,
$e^S$. The entropy $S$ of a space-time is given by the total of its
horizon areas, divided by four; it follows that $S = -2I^{\rm Re}$ in
the cosmological case~\cite{HawHor95}.  So Eq.~(\ref{eq-prob-measure})
above does indeed reflect the number of internal states.  If the black
hole space-time has lower entropy than the background, one obtains
$\Gamma<1$. Then the pair creation will be suppressed, as it should
be.

\subsection{Tunnelling Proposal} \label{ssec-tp}

The tunnelling proposal places different boundary conditions on the
wave function at small geometries in the Euclidean region.

The action~\rf{eq-action} is in general negative for a small boundary
geometry $ h_{ij} $. Thus $\Psi = e^{-I}$ is enhanced.
The proponents of the tunnelling proposal feel,
however, that the wave function ought to be suppressed in the
Euclidean region because it is supposed to be forbidden. They are
therefore forced to choose the
\begin{equation}
\Psi_{\rm {TP}} = \exp \left( +I \right)
\label{eq-tp}
\end{equation}
solution of the Wheeler-DeWitt equation as the boundary condition at
small $ h_{ij} $. This has the obvious disadvantage that it does not
reflect the entropy difference correctly. Transitions in the direction
of lower entropy are enhanced, rather than suppressed. This will lead
to absurd predictions in the context of pair creation.

In the following two sections we shall discuss the saddle-point
solutions needed to describe the pair creation of black holes on a
cosmological background~\cite{BouHaw95}. We shall use only the no
boundary proposal to calculate the probability measures and the pair
creation rate. The disastrous consequences of choosing the
prescription~\rf{eq-tp}, instead, will be discussed in
Sec.~\ref{sec-discussion}.

\sect{Instantons} \label{sec-fixed-cc}

We shall assume spherical symmetry.
Before we introduce a more realistic inflationary model, it is helpful
to consider a simpler situation with a fixed positive cosmological
constant $ \Lambda $ but no matter fields. We can then generalise
quite easily to the case where an effective cosmological ``constant''
arises from a scalar field.

\subsection{de~Sitter Space}

First we consider the case without black holes, a homogeneous
isotropic universe.  Since $\Lambda > 0$, its
spacelike sections will simply be round three-spheres. The wave
function is given by a path integral over all metrics on a
four-manifold $ M_+ $ bounded by a round three-sphere $ \Sigma $ of
radius $ a_{\Sigma} $.  The corresponding saddle-point solution is
the de~Sitter space-time.  Its Euclidean metric is that of a round
four-sphere of radius $\sqrt{3/\Lambda}$:
\begin{equation}
ds^2 = d\tau^2 + a(\tau)^2 d\Omega_3^2,
\end{equation}
where $\tau$ is Euclidean time, $d\Omega_3^2$ is the metric on the
round three-sphere of unit radius, and
\begin{equation}
a(\tau) = \sqrt{\frac{3}{\Lambda}} \sin \sqrt{\frac{\Lambda}{3}} \tau.
\label{eq-a-LS4}
\end{equation}

We can regard Eq.~\rf{eq-a-LS4} as a function on
the complex $\tau$-plane. On a line parallel to the imaginary
$\tau$-axis defined by $\tau^{\rm Re} = \sqrt{\frac{3}{\Lambda}} \,
\frac{\pi}{2} $, we have
\begin{equation} 
\left. a(\tau) \right|_{\tau^{\rm Re}=
                        \sqrt{\frac{3}{\Lambda}} \frac{\pi}{2}} = 
 \sqrt{\frac{3}{\Lambda}} \cosh \sqrt{\frac{\Lambda}{3}} \tau^{\rm Im}.
\end{equation}
This describes a Lorentzian de~Sitter hyperboloid, with $\tau^{\rm
  Im}$ serving as a Lorentzian time variable.  One can thus construct
a complex solution, which is the analytical continuation of the
Euclidean four-sphere metric.  It is obtained by choosing a contour in
the complex $\tau$-plane from $0$ to $\tau^{\rm Re} =
\sqrt{\frac{3}{\Lambda}} \, \frac{\pi}{2} $ (which describes half of
the Euclidean four-sphere) and then parallel to the imaginary
$\tau$-axis (which describes half the Lorentzian hyperboloid). The
geometry corresponding to this path is shown in (Fig.~1).
\begin{figure}[htb] 
   \hspace{.1\textwidth}
   \vbox{
    \epsfxsize=.8\textwidth
    \epsfbox{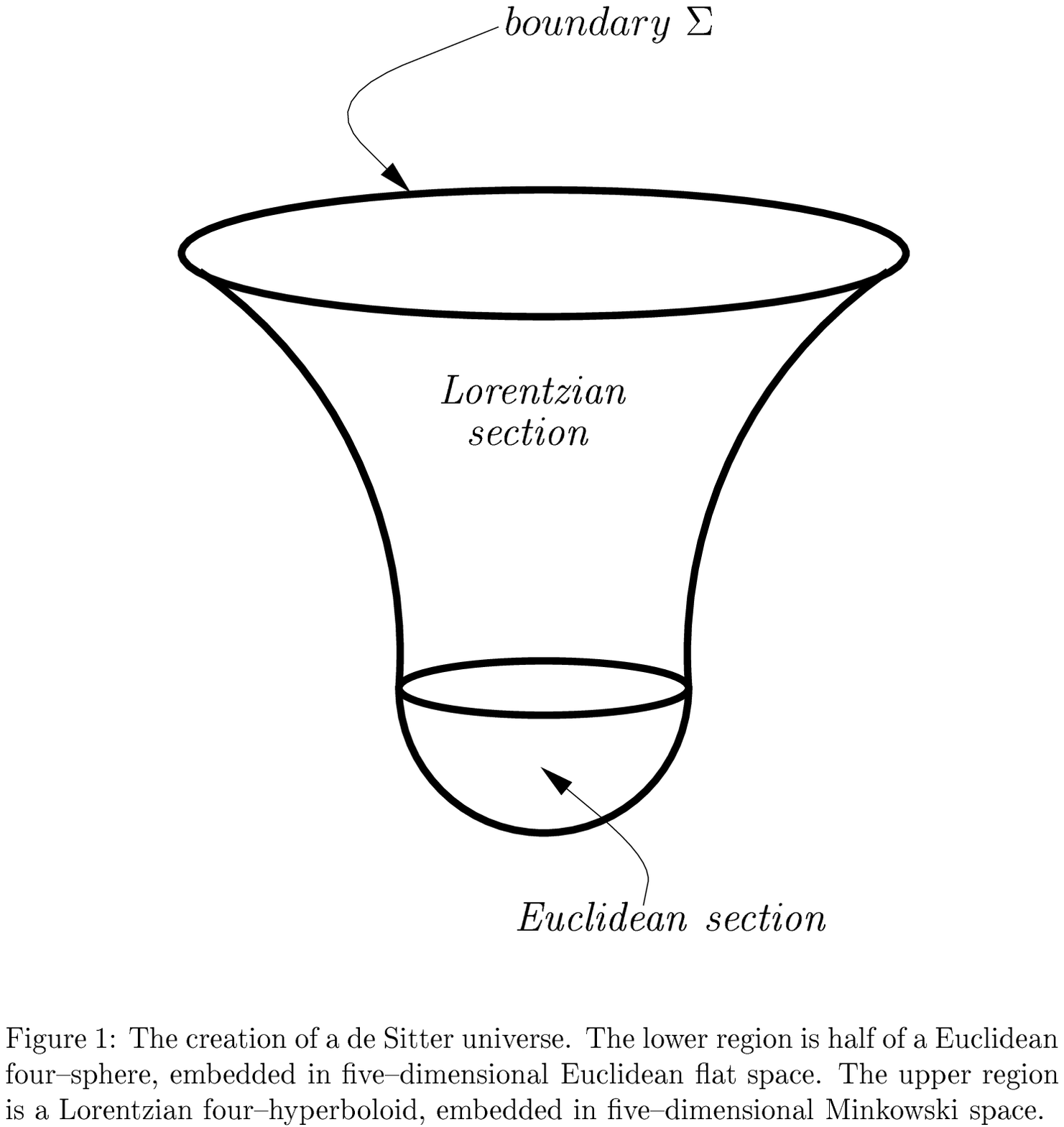}
    }
   \label{fig-creation}
\end{figure}

The Lorentzian part of the metric will contribute a purely imaginary
term to the action. This will affect the phase of the wave function
but not its amplitude. The real part of the action of this complex
saddle-point metric will be the action of the Euclidean
half-four-sphere:
\begin{equation}
I^{\rm Re}_{\rm de\, Sitter} = - \frac{3\pi}{2\Lambda}.
\end{equation}
Thus the magnitude of the
wave function will still be $ e^{3 \pi / 2 \Lambda} $, corresponding
to the probability measure
\begin{equation}
P_{\rm de\, Sitter} = \exp \left( \frac{3\pi}{\Lambda} \right).
\label{eq-prob-desitter}
\end{equation}

\subsection{Schwarzschild-de~Sitter Space}

We turn to the case of a universe containing a pair of black holes.
Now the cross sections $ \Sigma $ have topology $ S^2 \times S^1 $.
Generally, the radius of the $ S^2 $ varies along the $ S^1 $.  This
corresponds to the fact that the radius of a black hole immersed in
de~Sitter space can have any value between zero and the radius of the
cosmological horizon.  The minimal two-sphere corresponds to the
black hole horizon, the maximal two-sphere to the cosmological
horizon.  The saddle-point solution corresponding to this topology is
the Schwarzschild-de~Sitter universe.  However, the Euclidean section
of this spacetime typically has a conical singularity at one of its
two horizons and thus does not represent a regular
instanton~\cite{BouHaw96,GinPer83}. The only
regular Euclidean solution is the degenerate case where the black hole
has the maximum possible size. It is also known as the Nariai solution
and given by the topological product of two round two-spheres:
\begin{equation}
ds^2 = d\tau^2 + a(\tau)^2 dx^2 + b(\tau)^2 d\Omega_2^2,
\end{equation}
where $x$ is identified with period $2\pi$, $ d\Omega_2^2 = d\theta^2
+ \sin^2\! \theta\: d\varphi^2 $, and
\begin{equation}
a(\tau) = \sqrt{\frac{1}{\Lambda}} \sin \sqrt{\Lambda} \tau, \:
b(\tau) = \sqrt{\frac{1}{\Lambda}} = \mbox{const}.
\end{equation}

In this case the radius $b$ of the $ S^2 $ is constant in the $ S^1 $
direction. The black hole and the cosmological horizon have equal
radius and no conical singularities are present.  There will be no
saddle-point solution unless we specify $ b_{\Sigma}=1/\sqrt{\Lambda}
$.  Then the only variable we are free to choose on $\Sigma$ is the
radius $a_{\Sigma}$ of the one-sphere. In the Lorentzian section, the
one-sphere expands rapidly,
\begin{equation}
\left. a(\tau) \right|_{\tau^{\rm Re}=
                        \sqrt{\frac{1}{\Lambda}} \frac{\pi}{2}} = 
 \sqrt{\frac{1}{\Lambda}} \cosh \sqrt{\Lambda} \tau^{\rm Im},
\end{equation}
while the two-sphere (and, therefore, the black hole radius) remains
constant.  Again we can construct a complex saddle-point, which can be
regarded as half a Euclidean $ S^2 \times S^2 $ joined to half of the
Lorentzian solution. The real part of the action will be the action of
the half of a Euclidean $ S^2 \times S^2 $:
\begin{equation}
I^{\rm Re}_{\rm SdS} = -\frac{\pi}{\Lambda}.
\end{equation}
The corresponding probability measure is
\begin{equation}
P_{\rm SdS} = \exp \left( \frac{2\pi}{\Lambda} \right).
\label{eq-prob-nariai} 
\end{equation}
We divide this by the probability measure~\rf{eq-prob-desitter} for a
universe without black holes to obtain the pair creation rate of black
holes in de~Sitter space:
\begin{equation}
\Gamma =
\frac{P_{\rm SdS}}{P_{\rm de\, Sitter}} =
       \exp \left( - \frac{\pi}{\Lambda} \right).
\end{equation}
Thus the probability for pair creation is very low,
unless $ \Lambda $ is close to the Planck value, $ \Lambda = 1 $.

\subsection{Effective Cosmological Constant} \label{sec-effective-cc}

Of course the real universe does not have a large cosmological
constant. However, in inflationary cosmology it is assumed that the
universe starts out with a very large effective cosmological constant,
which arises from the potential $ V $ of a scalar field $ \phi $. The
exact form of the potential is not critical. So for simplicity we
chose $ V $ to be the potential of a field with mass $ m $, but the
results would be similar for a $ \lambda \phi^4 $ potential. To
account for the observed fluctuations in the microwave
background~\cite{COBE}, $ m $ has to be on the order of $ 10^{-5} $ to
$ 10^{-6} $~\cite{HalHaw85}.  The wave function $ \Psi $ will now
depend on the three-metric $ h_{ij} $ and the value of $ \phi $ on $
\Sigma $. For $ \phi > 1 $ the
potential acts like an effective cosmological constant
\begin{equation}
\Lambda_{\rm eff}(\phi) = 8 \pi V(\phi).
\end{equation}
There will again be complex saddle-points which can be regarded as a
Euclidean solution joined to a Lorentzian solution. Due to the time
dependence of $\Lambda_{\rm eff}$, however, one cannot find a path in
the $\tau$-plane along which the Euclidean and Lorentzian metrics will
be exactly real~\cite{BouHaw95}. Apart from this subtlety, the saddle
point solutions are similar to those for a fixed cosmological
constant, with the time-dependent $\Lambda_{\rm eff}$ replacing $
\Lambda $.  The radius of the pair created black holes will now be
given by $1/\sqrt{\Lambda_{\rm eff}}$.  As before, the magnitude of
the wave function comes from the real part of the action, which is
determined by the Euclidean part of the metric. This real part will be
\begin{equation}
I^{\rm Re}_{S^3} = - \frac{3 \pi}{2 \Lambda_{\rm eff}(\phi_0)}
\end{equation}
in the case without black holes, and
\begin{equation}
I^{\rm Re}_{S^2 \times S^1} = - \frac{\pi}{\Lambda_{\rm eff}(\phi_0)}
\end{equation}
in the case with a black hole pair. Here $ \phi_0 $ is the value of $
\phi $ in the initial Euclidean region. Thus the pair creation rate is
given by
\begin{equation}
\Gamma = 
\frac{P_{S^2 \times S^1}}{P_{S^3}} =
         \exp \left[ - \frac{\pi}{\Lambda_{\rm eff}(\phi_0)} \right].
\label{eq-suppression}
\end{equation}

\sect{Tunnelling vs. No Boundary Proposal} \label{sec-discussion}

In the previous sections we have used the no boundary proposal to
calculate the pair creation rate of black holes during inflation.  Let
us interpret the result, Eq.~\rf{eq-suppression}.  Since $ 0 <
\Lambda_{\rm eff} \leq 1 $, we get $\Gamma < 1$, and so black hole
pair creation is suppressed. In the early stages of inflation, when
$\Lambda_{\rm eff} \approx 1$, the suppression is week, and black
holes will be plentifully produced .  However, those black holes will
be very small, with a mass on the order of the Planck mass.  Larger
black holes, corresponding to lower values of $\Lambda_{\rm eff}$ at
later stages of inflation, are exponentially suppressed.  A detailed
analysis of their evolution~\cite{BouHaw96} shows that the small black
holes typically evaporate immediately, while sufficiently large ones
grow with the horizon and survive long after inflation ends.

We now understand how the standard prescription for pair creation,
Eq.~\rf{eq-pcr-usual}, arises from the no boundary proposal: By
Eq.~\rf{eq-prob-measure},
\begin{equation}
\Gamma = 
\frac{P_{S^2 \times S^1}}{P_{S^3}} =
\exp \left[ - \left( 2I^{\rm Re}_{S^2 \times S^1}-2I^{\rm Re}_{S^3}
  \right) \right],
\end{equation}
where $I^{\rm Re}$ denotes the real part of the Euclidean action of a
complex saddle-point solution. But we have seen that this real part is
equal to half of the action of the complete Euclidean solution. Thus
$I_{\rm obj} = 2I^{\rm Re}_{S^2 \times S^1}$ and $I_{\rm bg} = 2I^{\rm
  Re}_{S^3}$, and we recover Eq.~\rf{eq-pcr-usual}.

Let us return to the tunnelling proposal and see what results it would
have produced.  $\Psi_{\rm TP}$ is given by $e^{+I}$ rather than
$e^{-I}$.  This choice of sign is inconsistent with
Eq.~\rf{eq-pcr-usual}, as it leads to the inverse result for the pair
creation rate: $\Gamma_{\rm TP} = 1/\Gamma$.  In our case, we would
get $\Gamma_{\rm TP} = \exp ( + \pi/\Lambda_{\rm eff})$.  Thus black
hole pair creation would be enhanced, rather than suppressed.  This
means that de~Sitter space would decay: it would be catastrophically
unstable to the formation of black holes.  Since the radius of the
black holes is given by $1/\sqrt{\Lambda_{\rm eff}}$, the black holes
would be more likely the larger they were. Clearly, the tunnelling
proposal cannot be maintained.  On the other hand,
Eq.~\rf{eq-suppression}, which was obtained from the no boundary
proposal, is physically very reasonable.  It allows topological
fluctuations near the Planckian regime, but suppresses the formation
of large black holes at low energies.

We summarise.  The cosmological pair production of black holes
provides an ideal theoretical laboratory in which to examine the
question of the boundary conditions for the wave function of the
universe. The results could not be more decisive. The no boundary
proposal leads to physically sensible results, while the tunnelling
proposal predicts a disastrous enhancement of black hole production.

\end{document}